\documentclass[compsoc,conference,a4paper,10pt,times]{IEEEtran}
\IEEEoverridecommandlockouts
\makeatletter
\def\appendix{\par
  \setcounter{section}{0}%
  \setcounter{subsection}{0}%
  \gdef\thesection{\Alph{section}}}
\makeatother
\usepackage{cite}
\usepackage{amsmath,amssymb,amsfonts}
\usepackage{algorithmic}
\usepackage{graphicx}
\usepackage{textcomp}
\usepackage{bmpsize}
\usepackage{xcolor}
\usepackage{lipsum}
\usepackage[colorlinks=true,urlcolor=black]{hyperref}
\def\BibTeX{{\rm B\kern-.05em{\sc i\kern-.025em b}\kern-.08em
    T\kern-.1667em\lower.7ex\hbox{E}\kern-.125emX}}

\usepackage{pifont}
\usepackage{titlesec}
\usepackage{booktabs}

\usepackage{microtype} 
\usepackage{enumitem}

\newenvironment{tightitemize}{
  \begin{itemize}[noitemsep, topsep=0pt, leftmargin=*]
}{
  \end{itemize}
}

\newcommand{\cmark}{\textcolor{green}{\ding{51}}} 
\newcommand{\xmark}{\textcolor{red}{\ding{55}}} 
\def\BibTeX{{\rm B\kern-.05em{\sc i\kern-.025em b}\kern-.08em
    T\kern-.1667em\lower.7ex\hbox{E}\kern-.125emX}}
\newcommand{\debug}[0]{}
\ifdefined\debug
    \newcommand{\Jwd}[1]{\textcolor{orange}{Jack: #1}}
    \newcommand{\jdh}[1]{{\color{cyan} Jason: #1}}
    \newcommand{\Anh}[1]{\textcolor{cyan}{Anh: #1}}
    \newcommand{\Steven}[1]{\textcolor{green}{Steven: #1}}
    \newcommand{\Matthew}[1]{\textcolor{blue}{Matthew: #1}}
\else
    \newcommand{\Jwd}[1]{}
    \newcommand{\jdh}[1]{}
    \newcommand{\Anh}[1]{}
    \newcommand{\Steven}[1]{}
    \newcommand{\Matthew}[1]{}
\fi
\newcommand{\gulp}[1]{}
\begin{document}
\pagestyle{plain}
\pagenumbering{arabic} 

\title{PHASE: Passive Human Activity Simulation Evaluation}

\author{
\IEEEauthorblockN{Steven Lamp, Jason D. Hiser, Anh Nguyen-Tuong, Jack W. Davidson}
\IEEEauthorblockA{ 
\textit{University of Virginia}, Charlottesville, VA  USA \\
\{vxn3kr,hiser,an7s,jwd\}@virginia.edu}
}

\maketitle

\begin{abstract}

Cybersecurity simulation environments, such as cyber ranges, honeypots, and sandboxes, require realistic human behavior to be effective. To generate realistic activity, such simulation environments typically create synthetic user personas. However, no quantitative method exists to assess or improve the behavioral fidelity of synthetic user personas. To address this issue, this paper presents PHASE (Passive Human Activity Simulation Evaluation). PHASE operates by analyzing Zeek connection logs and scoring network activity behavior on a scale from human-predominant (human) to automation-predominant (non-human). Evaluated across eight diverse real-world datasets with established ground truth, PHASE's machine learning framework achieves on average 91\% accuracy in network activity classification. PHASE operates non-intrusively through passive network log analysis, resulting in no visible signs of surveillance. Furthermore, PHASE is augmented with SHAP (SHapley Additive exPlanations) analysis to uncover temporal and behavioral signatures indicative of genuine human users. Our analysis reveals a critical insight: when an activity occurs is as important as what activity occurs for behavioral fidelity. Human behavior exhibits natural temporal gaps corresponding to meetings, breaks, and task transitions that distinguish it from automated systems. We further evaluate a widely used synthetic user persona tool and develop an enhanced configuration utilizing temporal camouflage strategies, achieving up to 55 percentage points of improvement in behavioral fidelity scores. Critically, PHASE reveals that human behavior is highly contextually variable, and strategies effective in structured environments may fail in contexts with fluid behavioral patterns, demonstrating PHASE's essential role in revealing gaps between expected and actual authenticity. To facilitate reproducibility and enable further research, the PHASE framework, models, and datasets will be made available to the research community.

\end{abstract}


\section{Introduction}
Simulation environments are central to modern cybersecurity research, enabling controlled experimentation, adversarial testing, and the development of automated defense strategies. For these environments to provide meaningful insights, they must faithfully recreate the complex network behaviors seen in real-world settings~\cite{Blythe_Botello_Sutton_Mazzocco_Lin_Spraragen_Zyda_2011}. Real behaviors include background system traffic and the dynamic activities of human users. Without realistic human network activity, simulations risk being unrepresentative of the real world, which may lead to poor performance once said defenses are deployed.

However, it is often infeasible or unethical to include real users in large-scale simulations.  Instead, most environments rely on \emph{Synthetic User Personas} (SUPs) to emulate human behavior. SUPs such as the widely used MITRE Caldera Human Plugin (MCHP) are embedded into large research efforts, including DARPA CHASE and DARPA CASTLE~\cite{CHASE, CASTLE}. These projects used SUPs to generate user-like traffic to support simulations of office, campus, or other enterprise networks. 

\begin{figure}[hbtp]
    \includegraphics[width=\columnwidth]{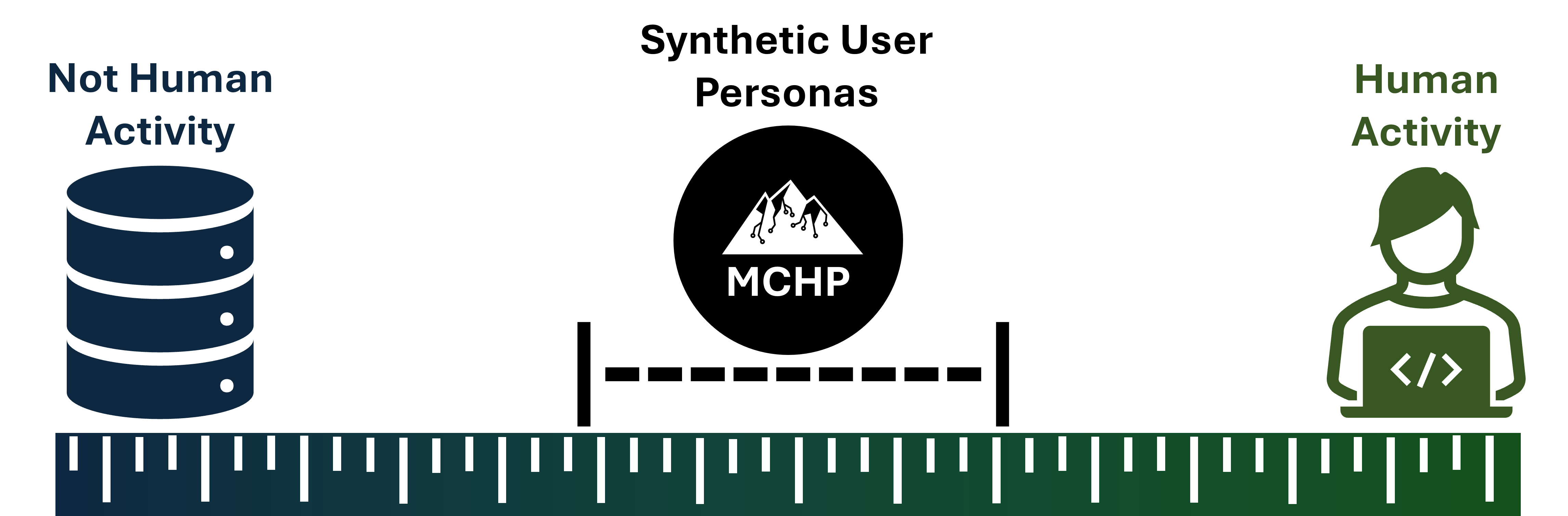}
    \caption{Measuring stick of network activity fidelity}
   \label{fig:teaser}
\end{figure}

Despite the utility of these SUP systems, currently there is no standardized quantitative framework for evaluating how closely SUP-generated activity matches that of real human users, like a measuring stick, as illustrated in~\autoref{fig:teaser}. This gap represents a significant obstacle across all cybersecurity simulation environments, whether enterprise network testbeds, honeypots, cloud-based sandboxes, or cyber defense training platforms. Without robust measures of behavioral fidelity, researchers have little guidance in tuning or improving SUP behavior, no consistent way to compare different systems or configurations, and cannot validate that their environments accurately represent real operational conditions.

To address this evaluation gap, we present \emph{PHASE} (\underline{P}assive \underline{H}uman \underline{A}ctivity \underline{S}imulation \underline{E}valuation), a machine learning based measurement and analysis framework for assessing the realism of simulated user activity. PHASE operates entirely passively, analyzing raw network telemetry to score network activity patterns on a scale from human-predominant (human) to automation-predominant (non-human). This passive design ensures that PHASE does not introduce artificial artifacts, thereby preserving the integrity of the simulation environment. The framework employs deep neural networks trained on labeled network activity to generate a score that represents the SUP's fidelity. The PHASE framework also creates SHAP (SHapley Additive exPlanations) analysis graphs, which can be used to identify behavioral gaps, refine simulation models, or evaluate new SUP designs.

PHASE achieves an average accuracy of 91\% in distinguishing human from automated network activity in eight temporally and environmentally distinct datasets. The framework's interpretability features, powered by SHAP analysis, reveal specific temporal and behavioral signatures that characterize genuine human activity, enabling data-driven improvements to synthetic personas rather than relying on subjective assessments.

PHASE's interpretability analysis reveals a critical insight for synthetic persona design: for behavioral fidelity, when an activity occurs is as important as what activity occurs. Human network behavior exhibits natural temporal gaps corresponding to meetings, breaks, and task transitions that distinguish it from automated systems. Using these temporal patterns through strategic idle periods, synthetic personas can achieve dramatic improvements in behavioral realism, with our evaluation demonstrating up to 55 percentage points of improvement. However, PHASE also reveals that human behavior is highly contextually variable. Strategies that appear human-like in structured operational environments (such as academic fall/spring semesters with rigid schedules) may fail in contexts with more fluid behavioral patterns (such as summer periods with flexible schedules). This contextual sensitivity demonstrates the essential role of PHASE in revealing the gaps between expected realism and actual authenticity across diverse operational environments.


\noindent This paper addresses four research questions:
\begin{tightitemize}
    \item \textbf{RQ1:} Can passive network log analysis reliably 
    distinguish human-predominant from automation-predominant network 
    activity across diverse operational environments? 

    \item \textbf{RQ2:} Which temporal and behavioral feature signatures are most indicative of genuine human network activity, and how do these signatures vary across operational contexts?
    
    \item \textbf{RQ3:} Do existing synthetic user persona tools, 
    specifically the widely deployed MITRE Caldera Human Plugin, produce 
    network activity that is behaviorally indistinguishable from real 
    human users? 
    
    \item \textbf{RQ4:} Can data-driven insights derived from behavioral analysis be used to measurably improve the humanness of synthetic user personas?
\end{tightitemize}

\noindent To address these research questions, this paper makes the 
following contributions:
\begin{tightitemize}
    \item A novel measurement and analysis framework that scores synthetic user persona behavioral fidelity by classifying network activity patterns as human-predominant or automation-predominant. 
    
    \item A SHAP-based interpretability analysis that reveals specific temporal and behavioral signatures indicative of genuine human network activity, demonstrating that \textbf{when} activity occurs is critically important for behavioral fidelity, enabling systematic improvements to synthetic personas.

    \item An evaluation and enhancement of the widely deployed MITRE Caldera Human Plugin, demonstrating measurable improvements in behavioral realism through data-driven configuration adjustments.

    \item A novel data processing pipeline to take raw connection logs and transform them for time series machine learning analysis. 
    
    \item Four temporally distinct datasets collected over the course of four semesters, each encompassing several contiguous months of activity.
    
\end{tightitemize}

This paper first presents background information necessary for understanding the PHASE framework (\autoref{sec:background}), then details the PHASE DNN architecture and feature engineering (\autoref{section:phase-design}), followed by a description of the datasets used for evaluation (\autoref{section:phase-evaluation}), a performance evaluation on real-world network traces (\autoref{subsection:phase-model-training}), and practical validation through a widely used synthetic human plugin case study (\autoref{section:sup-methodology}). We then discuss related work (\autoref{section:related-work}), limitations and future directions (\autoref{section:future-work-ext}), and ethical considerations (\autoref{section:ethical-considerations}), before concluding (\autoref{section:conclusion}).


\begin{figure}[hbtp]
\centering
 \includegraphics[width=1\columnwidth]{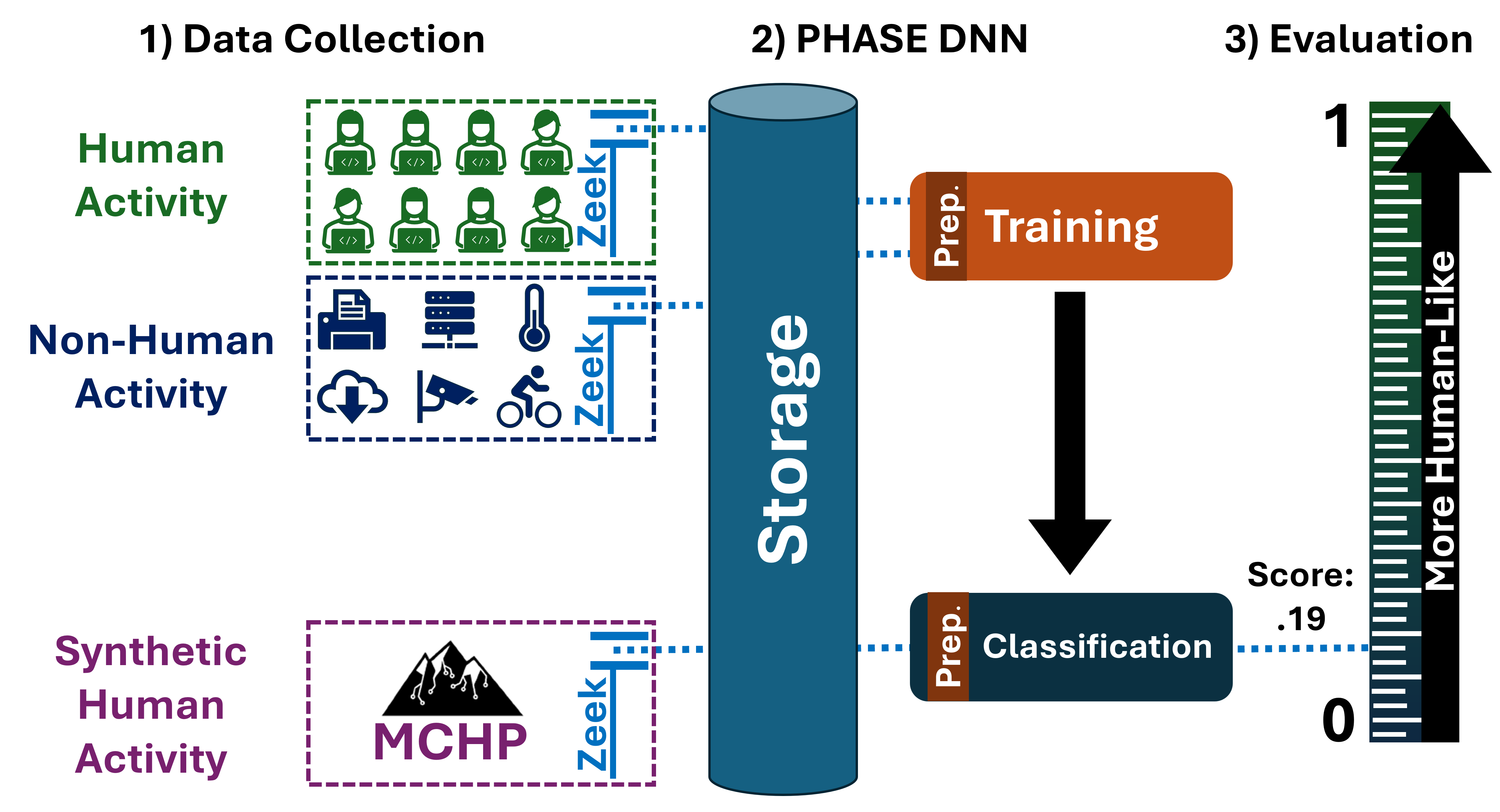}
  \caption{PHASE Framework for SUP evaluation}
  \label{fig:framework}
\end{figure}

\section{Background}
\label{sec:background}

This section introduces the core technologies underlying the PHASE framework. PHASE relies on three foundational components: passive network telemetry for observing device behavior, machine learning interpretability techniques for understanding classification decisions, and synthetic user persona systems used in cybersecurity simulation environments.

Zeek is a passive, open-source network traffic analyzer widely deployed in cybersecurity operations and research~\cite{zeek}. Rather than storing raw packet captures, Zeek processes live network traffic and produces structured connection logs that record metadata for every observed connection. These logs capture information such as IP addresses, ports, protocol, byte counts, and connection state, without storing payload content. Zeek is a widely-used open-source tool for network security monitoring, and its widespread adoption ensures that PHASE can be deployed across diverse network infrastructures without requiring specialized data collection mechanisms.

SHAP (SHapley Additive exPlanations) is a framework for interpreting machine learning model predictions by quantifying the contribution of each input feature to a given output~\cite{shapley}. In practical terms, SHAP answers the question: given that the model classified this device as human, which specific features and at what times of day drove that decision? Rather than treating the model as a black box, SHAP produces explainable, feature-level evidence that security practitioners can reason about directly. In PHASE, SHAP analysis is applied to reveal which temporal and behavioral features most strongly drive human versus non-human classification decisions, enabling data-driven improvements to synthetic user persona design rather than relying on subjective assessment.

The MITRE Caldera Human Plugin (MCHP) is a widely deployed synthetic user persona tool used to simulate realistic human network activity in cybersecurity simulation environments, and was originally released as a part of the MITRE Caldera Cyber Range~\cite{alford2022caldera}. MCHP provides customizable workflows that emulate common user behaviors including web browsing and shell command execution. It has been embedded in large-scale research efforts including DARPA CHASE and DARPA CASTLE~\cite{CHASE, CASTLE}, making it a representative and practically relevant target for behavioral fidelity evaluation.
\section{PHASE Design}
\label{section:phase-design}
The core of PHASE is a deep neural network that classifies device activity patterns as human-predominant or automation-predominant. The framework operates on passively collected Zeek connection logs to detect temporal and behavioral patterns indicative of genuine user activity~\cite{zeek}.

PHASE is designed to utilize connection logs as its primary data source due to their unique advantages for behavioral analysis in operational network environments. Connection logs are widely generated through the Zeek network activity monitor, which has become a standard tool in cybersecurity operations and research~\cite{zeek}. This widespread adoption ensures that PHASE can be easily deployed across diverse network infrastructures without requiring specialized data collection mechanisms.

The passive nature of connection log collection is particularly advantageous for behavioral fidelity analysis. Unlike active monitoring techniques that inject probe traffic or modify network behavior, connection logs capture genuine user activity without introducing artifacts that could contaminate behavioral patterns. This passivity ensures that the observed behaviors authentically represent real user and device actions, rather than responses to the monitoring infrastructure.

Connection logs provide high-resolution temporal information with precise timestamps for each network event~\cite{zeek}. This temporal granularity enables PHASE to analyze behavioral patterns over time, capturing the rhythms and sequences characteristic of human versus automated activity. As discussed in~\autoref{subsection:SHAP-time-wise}, temporal behavioral analysis is essential to distinguish between human users, who exhibit irregular patterns influenced by daily rhythms and work schedules, and automated systems, which typically demonstrate consistent, repetitive behaviors.~\autoref{tab:connection_log} presents the connection log features utilized by PHASE. These features were selected because they contain dynamic information that changes based on the characteristics of each connection. 

\begin{table}[h]
\setlength{\tabcolsep}{3pt}          
\centering
\resizebox{\columnwidth}{!}{%
    \begin{tabular}{ll}
        \toprule
        \textbf{Feature} & \textbf{Description} \\
        \midrule
        Timestamp         & Time the connection was observed \\
        local\_orig       & Indicator of if the source is local \\
        local\_resp       & Indicator of if the destination is local \\
        conn\_state       & State of the connection (e.g., ESTABLISHED) \\
        proto             & Protocol used (e.g., TCP, UDP) \\
        service           & Service used in the connection \\
        duration          & Duration of the connection \\
        history           & History of the connection \\
        missed\_bytes     & Number of bytes missed in the connection \\
        id.orig\_p        & Port number of the source \\
        id.resp\_p        & Port number of the destination \\
        orig\_bytes       & Number of bytes sent from the source \\
        resp\_bytes       & Number of bytes received by the destination \\
        orig\_ip\_bytes   & Number of IP bytes sent from the source \\
        resp\_ip\_bytes   & Number of IP bytes received by the destination \\
        orig\_pkts        & Number of packets sent from the source \\
        resp\_pkts        & Number of packets received by the destination \\
        norm\_vol         & Volume of the connections collapsed per minute \\
        \bottomrule
    \end{tabular}
}
\vspace{2 pt}
\caption{Connection Log Features}
\label{tab:connection_log}
\end{table}

\begin{figure}[hbtp]
  \centering
  \includegraphics[width=\columnwidth]{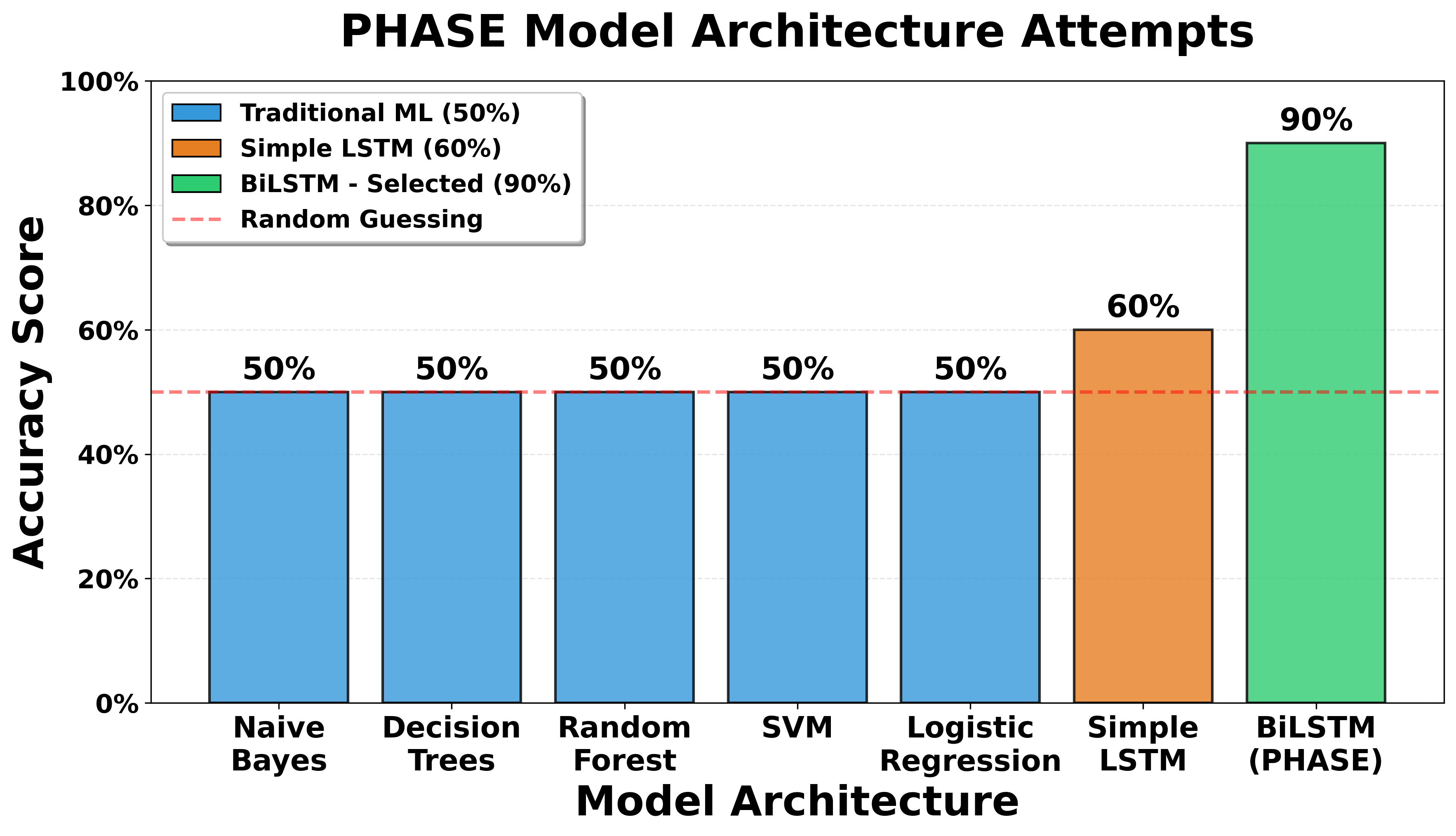}
  \caption{PHASE initial attempts before DNNs}
  \label{fig:phaseinit}
\end{figure}

Our initial attempts to classify connection logs treated each connection as an independent observation, applying traditional machine learning techniques such as decision trees and Naive Bayes to individual connection features. These approaches resulted in poor performance (\autoref{fig:phaseinit}), achieving approximately 50\% accuracy. The fundamental limitation was clear: these models could not capture the temporal dependencies inherent in human behavior, where activity patterns unfold over minutes and hours rather than within isolated connections.

Recognizing that human network behavior exhibits temporal structure, we shifted to time series modeling using basic LSTM architectures. This transition represented a conceptual change: rather than classifying individual connections, we began analyzing sequences of aggregated minute-level activity throughout a day. Early LSTM models showed modest improvement (approximately 60\% accuracy), validating that temporal patterns are indeed discriminative. However, these initial sequential models remained insufficient for robust classification, struggling to capture both short-term behavioral transitions (e.g., starting a task) and long-term daily rhythms (e.g., work hours versus idle periods).

PHASE therefore adopts a more sophisticated time series-based approach inspired by Human Activity Recognition (HAR) models (\autoref{section:related-work}). Human network behavior exhibits temporal patterns at multiple scales: daily rhythms, work patterns, and activity bursts that create dependencies across time. The hybrid CNN-BiLSTM-Attention architecture (\autoref{subsection:DNNLayerComp}) was specifically designed to address these multi-scale temporal dependencies. The convolutional layers capture short-term patterns within minute-level windows, the bidirectional LSTM layers model long-range dependencies across the full day, and attention mechanisms highlight the most behaviorally significant time periods. This architecture treats each day as a sequence of 1,440 one-minute intervals, enabling the model to learn how activity at each time step relates to past and future behavior throughout the entire daily cycle.


This temporal classification approach is implemented through three components: a hybrid CNN-BiLSTM-Attention architecture that captures both short and long-range dependencies (\autoref{subsection:DNNLayerComp}), a preprocessing pipeline that transforms raw network telemetry into structured time series traces (\autoref{subsection:pprocessing}), and supervised training on labeled real-world data to distinguish human from automated activity (\autoref{subsection:Training}).

\subsection{DNN Layer Composition}
\label{subsection:DNNLayerComp}
\begin{figure}[hbtp]
\centering
    \includegraphics[width=.9\columnwidth]{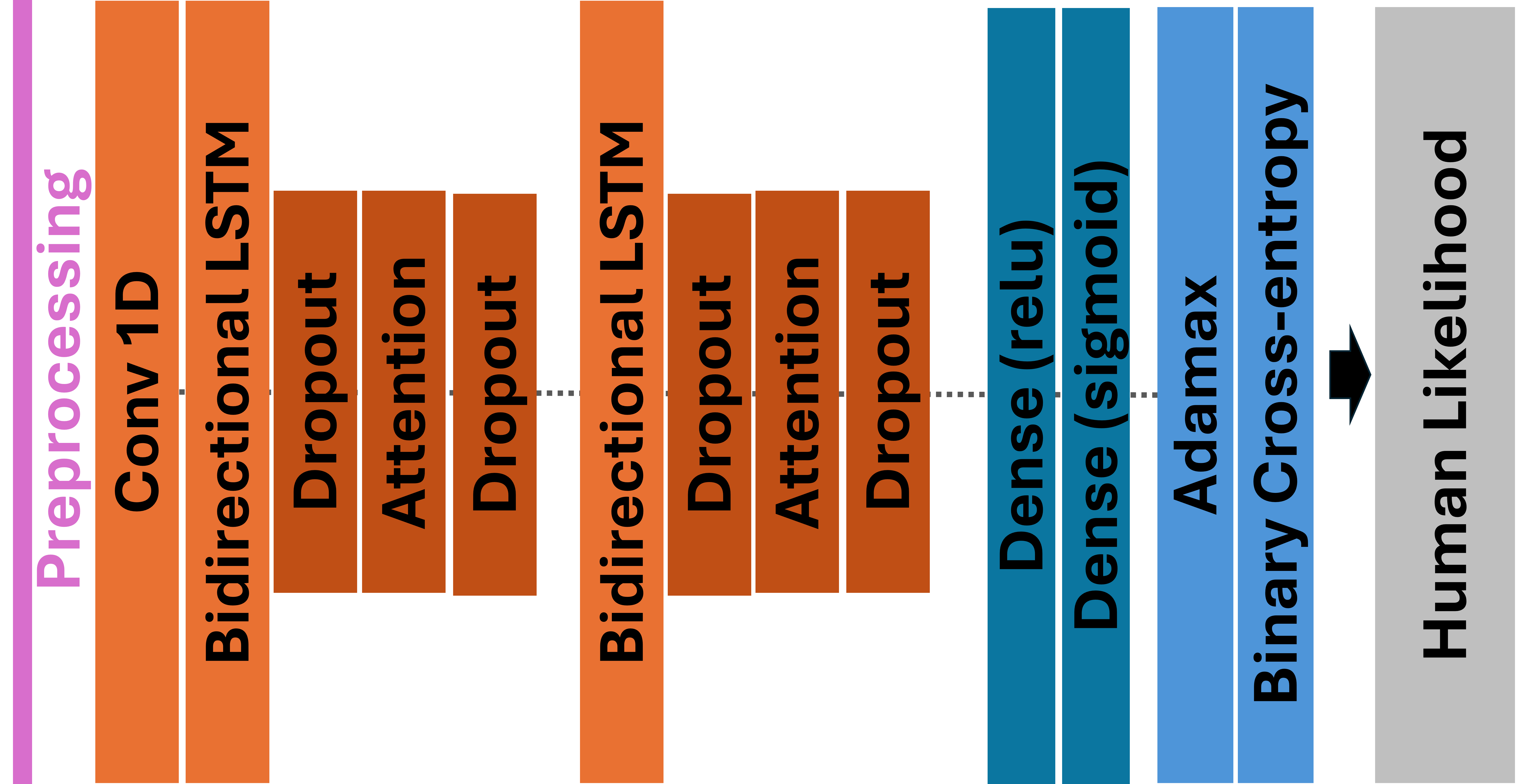} 
    \caption{Block diagram of the PHASE DNN using a hybrid LSTM architecture} 
    \label{fig:modelArch} 
\end{figure}

As shown in \autoref{fig:modelArch}, the PHASE architecture comprises three layers that capture different temporal aspects of network behavior. A 1D convolutional layer identifies short-term patterns within the minute-level data~\cite{keras}. Two bidirectional LSTM layers capture long-term dependencies by processing sequences of data in both directions~\cite{LSTM, bilstm}. Multi-headed attention mechanisms highlight the most relevant time steps for classification decisions~\cite{vaswani2023attentionneed}. Dropout (0.2) and Adamax optimization provide regularization and training stability~\cite{kingma2017adammethodstochasticoptimization}. This hybrid architecture draws inspiration from human activity recognition approaches that successfully capture multi-scale temporal dependencies, as detailed in~\autoref{section:related-work}.

\subsection{PHASE Network Log Preprocessing}
\label{subsection:pprocessing}
To use Zeek connection logs for machine learning time series analysis in the PHASE Framework, we need to unify the uneven time periods at which connection logs were collected, convert text categories to numbers, and normalize all data to prevent any data leakage or bias during training.

\subsubsection{Temporal Alignment}
\label{subsubsection:uneventime}
\begin{figure}[hbtp]
    \centering
    \includegraphics[width=.9\columnwidth]{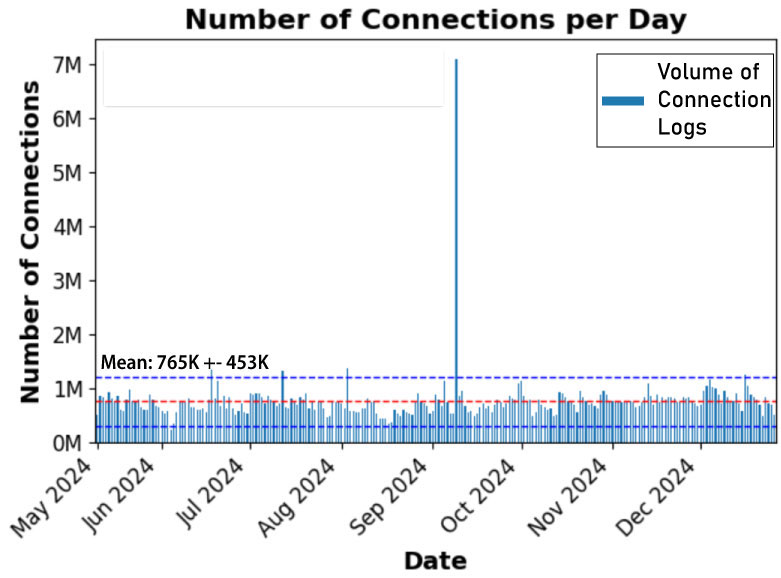} 
    \caption{The extreme peak represents students adjusting classes before the semester's drop/add deadline for classes, illustrating the natural variability in real-world network data.} 
    \label{fig:numConPerDay} 
\end{figure}

Zeek connection logs are recorded at irregular intervals, creating bursty and variable data patterns as shown in~\autoref{fig:numConPerDay}. 

To enable temporal modeling, each device's daily activity was divided into 1,440 one-minute intervals. Within each minute, connection logs were aggregated into a single record. Quantitative features were averaged, while categorical features used the most frequent value, known as modal encoding~\cite{pedregosa2011scikit}. Minute-level granularity was selected after testing coarser intervals, providing a balance between temporal detail and training efficiency. IP addresses were used only for data organization, not as model inputs, ensuring the model learns general behavioral patterns rather than device-specific traits.


String-based features (connection states, protocol types) were converted to numerical form using Scikit-learn's LabelEncoder~\cite{pedregosa2011scikit}. Port numbers were treated as categorical rather than continuous variables, since computing average port values (e.g., 223.5) lacks semantic meaning. Missing values were filled with -1 to remain outside the encoding range while maintaining compatibility with subsequent processing.

\subsubsection{Feature Scaling and Consistency}
\label{subsubsection:preventbias}
All numerical attributes were normalized using Scikit-learn's MinMaxScaler package to ensure consistent feature scaling~\cite{pedregosa2011scikit}. The fitted encoders and scalers were saved for reuse during inference, preventing data shift and maintaining stable model performance across different deployment scenarios.

\subsection{Model Training}
\label{subsection:Training}
The PHASE model was trained to classify network activity as automation-predominant (class 0) or human-predominant (class 1) network activity using the datasets described in~\autoref{subsection:cs-dept-datasets}. Real-world network datasets naturally exhibit class imbalance, which can bias model predictions toward the majority class. To mitigate this effect, we employed stratified 10-fold cross-validation to preserve class distributions across folds.

To prevent data leakage, the datasets were split into train, test, and validation datasets for each fold. Within each fold, the majority class (usually automation-predominant activity) was undersampled during training to balance class representation, ensuring the model does not learn to guess a single class~\cite{lemaitre2017imbalanced}. The model was trained for up to 1,000 epochs using the Adamax optimizer (learning rate 0.001), with early stopping implemented to halt training if the training loss failed to improve by 0.001 over 20 consecutive epochs. The framework automatically restored the best-performing weights observed during training. A random dataset validation test was performed to verify pipeline integrity and confirm the absence of data leakage, as detailed in~\autoref{app:data-leakage}.

\section{Datasets}
\label{section:phase-evaluation}


To comprehensively evaluate PHASE across diverse operational environments, we assembled eight datasets: two publicly available datasets from the Collegiate Penetration Testing Competition (CPTC) global finals in 2023 and 2024~\cite{cptc8_dataset_2023, cptc9_dataset_2024}, two datasets from a full academic network known as the PCORE datasets~\cite{SENTINEL}, and four datasets from a computer science department network with DNS-verified labels spanning up to 125 days. This multi-source approach enables rigorous assessment of PHASE's generalization across different network scales, operational contexts, and temporal periods.


\subsection{CPTC Datasets}
\label{subsubsection:CPTC-construction}
The CPTC 2023 and 2024 datasets are publicly available network captures from the Collegiate Penetration Testing Competition (CPTC) global finals. CPTC is the largest offense-based collegiate cybersecurity competition, with hundreds of students competing annually for spots on the 12-15 teams that advance to the finals. CPTC 2023 simulated a hotel and hospitality company network over 8 days with 13 human competitor sources and 19 automated services simulating enterprise infrastructure, while CPTC 2024 simulated an air and rail transportation company over 4 days with 14 human competitor sources and 23 automated services~\cite{cptc8_dataset_2023, cptc9_dataset_2024}.

The CPTC datasets present a unique challenge: multiple competitors may be logged into a single device simultaneously, meaning one network source can exhibit diverse human behavioral patterns from different individuals. Ground truth labels for each IP address were provided by competition organizers, distinguishing human competitor sources from automated enterprise services. These datasets offer a unique evaluation context distinct from operational academic networks, representing security-focused scenarios with intentional adversarial activity where multiple human actors generate traffic from shared infrastructure.

The CPTC datasets were originally collected through combinations of Splunk stream and netflow data sources~\cite{splunk_stream2025}. To enable PHASE evaluation, we converted this heterogeneous data into standardized Zeek connection log format through a multi-step process: extracting connection metadata from Splunk journals, merging with netflow data where available (CPTC 2023) or reconstructing connection states from application-layer indicators (CPTC 2024), and formatting the enriched records as Zeek logs compatible with PHASE's preprocessing pipeline. \autoref{table:cptc_data_availability} details the features available in each dataset after conversion.

\begin{table}[h!]
\centering
\small
\renewcommand{\arraystretch}{0.85}
\hspace{-0.1in}
\resizebox{1.02\columnwidth}{!}{%
\setlength{\tabcolsep}{0pt}
\begin{tabular}{lcc}
\toprule
\textbf{Feature} & \textbf{CPTC 2024} & \textbf{CPTC 2023} \\
                 & {Stream Only} & {Stream+Netflow} \\
\midrule
Timestamp (\texttt{ts}) & \cmark & \cmark \\
Flow ID (\texttt{uid}) & \cmark & \cmark \\
Source IP (\texttt{id.orig\_h}) & \cmark & \cmark \\
Source Port (\texttt{id.orig\_p}) & \cmark & \cmark \\
Destination IP (\texttt{id.resp\_h}) & \cmark & \cmark \\
Destination Port (\texttt{id.resp\_p}) & \cmark & \cmark \\
Protocol (\texttt{proto}) & \cmark & \cmark \\
Service (\texttt{service}) & \cmark & \cmark \\
Duration (\texttt{duration}) & \cmark & \cmark \\
Originator Bytes (\texttt{orig\_bytes}) & \cmark & \cmark \\
Responder Bytes (\texttt{resp\_bytes}) & \cmark & \cmark \\
Connection State (\texttt{conn\_state}) & Reconstructed & \xmark \\
Local Originator (\texttt{local\_orig}) & Reconstructed & \xmark \\
Local Responder (\texttt{local\_resp}) & Reconstructed & \xmark \\
Missed Bytes (\texttt{missed\_bytes}) & \xmark & \xmark \\
TCP History (\texttt{history}) & \xmark & \cmark \\
Originator Packets (\texttt{orig\_pkts}) & \xmark & \cmark \\
Originator IP Bytes (\texttt{orig\_ip\_bytes}) & \xmark & \cmark \\
Responder Packets (\texttt{resp\_pkts}) & \xmark & \cmark \\
Responder IP Bytes (\texttt{resp\_ip\_bytes}) & \xmark & \cmark \\
Tunnel Parents (\texttt{tunnel\_parents}) & \xmark & \xmark \\
\bottomrule
\end{tabular}
}
\vspace{2pt}
\caption{Data Availability by CPTC Dataset}
\label{table:cptc_data_availability}
\end{table}

\subsection{PCORE Datasets}
\label{subsubsection:PCORE-construction}
The PCORE academic university datasets were acquired from Virginia Tech's university network environment spanning beyond a single academic department. Due to the large scale of network activity, we extracted two datasets from the same 8-day period to provide different evaluation contexts: a small dataset with 665 IP addresses (332 human and 333 non-human) capturing 29.1 million connection logs, and a large dataset with 5,560 IP addresses (2,780 human and 2,780 non-human) capturing 260.1 million connection logs.

These datasets reflect genuine operational network behavior from a diverse university population including students, faculty, staff, and institutional services across multiple departments and functions. The datasets offer balanced class distributions at significantly larger scales than the CPTC datasets, though the 8-day observation period limits assessment of long-term behavioral patterns. Unlike the CPTC datasets with organizer-provided information about human-operated IPs, no explicit information about ground truth labels is explicitly available  for this operational university network. \autoref{subsection:creatinglabels} describes how we created labels for this dataset.

\subsection{CS Dept. Datasets}
\label{subsection:cs-dept-datasets}
The CS Dept. datasets provide comprehensive evaluation coverage across both temporal and scale dimensions. We constructed these datasets by deploying a dedicated collection system at an operational computer science department network at the University of Virginia with over 1,530 endpoints, including students, professors, and staff. This dataset addresses the limitations in both the CPTC and PCORE datasets. While the CPTC datasets offer well-labeled competitive scenarios, they are limited by brief duration (4-8 days) and small device populations (13-23 IP addresses). The PCORE datasets, though containing thousands of IP addresses, lack verified ground truth labels or DNS information. In contrast, the CS Dept. datasets span 96 contiguous days on average (with a range of 52 to 125 days) with 100-250 devices per semester, providing both extended temporal coverage and DNS information from which we can create ground truth labels (Section~\ref{subsection:creatinglabels}), which is necessary for capturing authentic behavioral patterns across different academic cycles and user populations. The CS Dept. datasets span four consecutive semesters, providing over a year of continuous data. These datasets provide the most reliable foundation for evaluating PHASE's ability to distinguish human-predominant from automation-predominant network activity across changing behavioral patterns.

\begin{figure}[hbtp]
  \centering
  \includegraphics[width=.9\columnwidth]{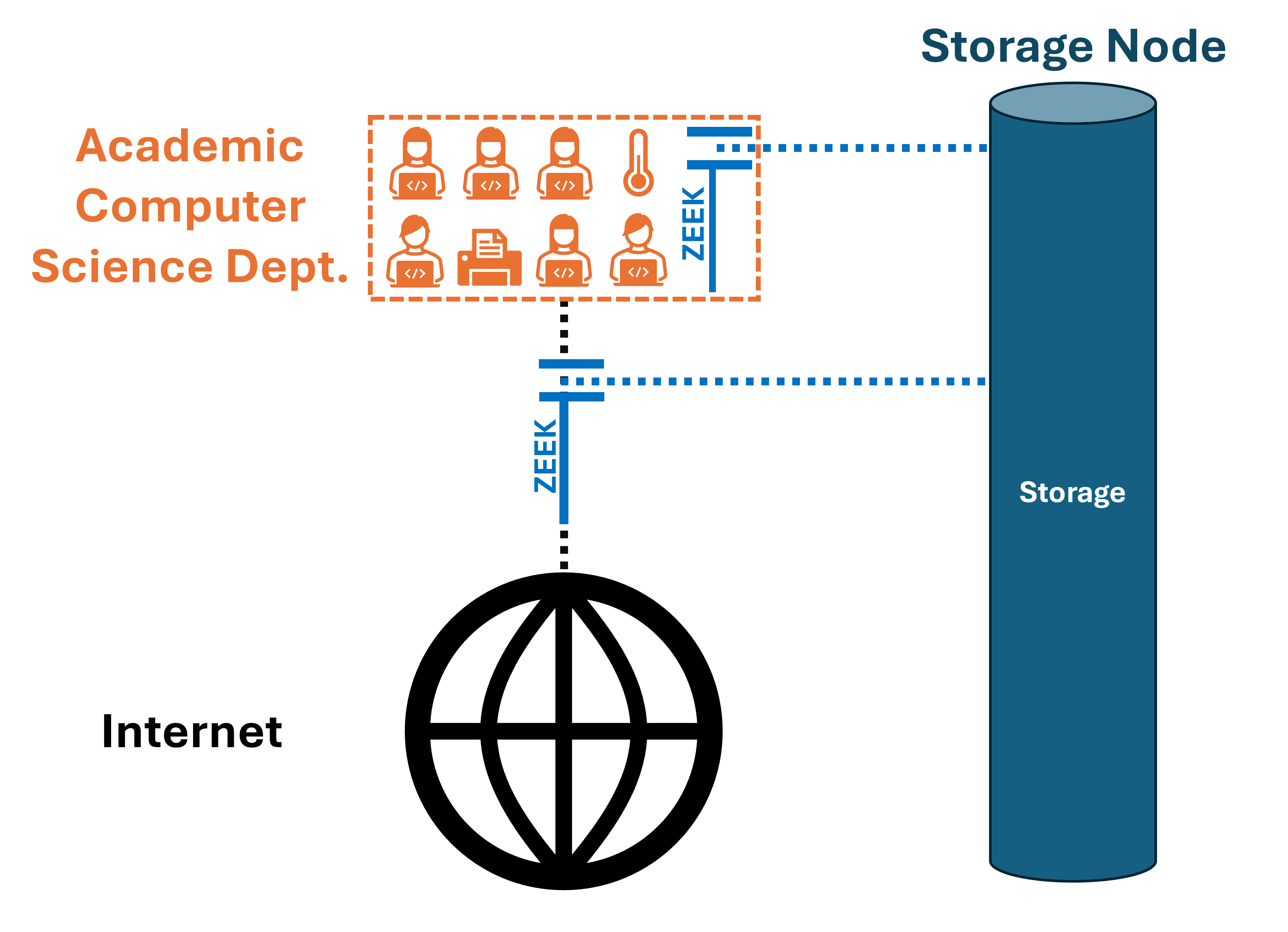}
  \caption{PHASE data collection using SENTINEL architecture}
  \label{fig:dataCollection}
\end{figure}

As shown in~\autoref{fig:dataCollection}, the dataset collection system used an architecture similar to prior work, where network traffic was mirrored to an internal server cluster equipped with Zeek sensors for monitoring~\cite{zeek, SENTINEL}. This server then digested the raw network traces into connection logs. All raw network traces are destroyed, and the only data at rest is the resulting connection logs. Data collection and privacy practices are detailed further in~\autoref{section:ethical-considerations}.

\noindent Network traffic collection spanned four consecutive academic semesters:
\begin{tightitemize}
    \item Summer 2024: June 17 to August 8
    \item Fall 2024: August 12 to December 16
    \item Spring 2025: January 13 to May 9
    \item Summer 2025: May 18 to August 9
\end{tightitemize}


Network traffic was collected by deploying Zeek sensors on a mirrored network port within the computer science department, capturing all traffic traversing the department network boundary. The monitored population includes a diverse set of endpoint types: student and faculty workstations, personal laptops, research servers, web servers, and IoT devices such as building automation systems. Zeek operates as a passive network analyzer, meaning it observes traffic without interacting with or modifying it. Zeek records connection-level metadata for all traffic on the wire, regardless of protocol or application. Payload content is never accessible or stored at any stage of the pipeline. Monitoring was scoped to the computer science department network for the CS Dept. datasets, while the CPTC and PCORE datasets were collected by third parties under different network scopes as described in~\autoref{subsubsection:CPTC-construction} and~\autoref{subsubsection:PCORE-construction} respectively.

Each academic period was collected separately to ensure consistent device-to-label mappings throughout the observation window. As students, faculty, and staff cycled through the network each semester and infrastructure upgrades occurred during inter-semester breaks, IP-to-device associations changed, requiring new ground truth labeling for each period. This semester-based segmentation preserves labeling accuracy while capturing significant variation in activity patterns due to changing schedules, user populations, and network configurations. The summer semesters differ most heavily from the fall and spring semesters, because class schedules are compressed, work hours in academia tend to be more varied, and people are more likely to travel for vacations and conferences. The collected connection log features (\autoref{tab:connection_log}) contain no personally identifiable information except IP addresses, which were anonymized per IRB protocols (\autoref{section:ethical-considerations}).

The four datasets each comprise over 150 GB of network activity, collected over 52, 125, 116, and 89 days, respectively.  The datasets are collected from an operational network of over 1,530 endpoints.



\subsection{Creating Ground-Truth Labels}
\label{subsection:creatinglabels}

\subsubsection{Labeling CPTC} 
Supervised training requires explicit device-level labels distinguishing human-predominant (human) from automation-predominant (non-human) activity to enable supervised training. 
For the CPTC datasets, labeling is straightforward: competition organizers recorded which IP addresses corresponded to human competitors, allowing us to manually annotate each device based on its known role.

\subsubsection{Labeling CS Dept.}
Unlike CPTC, the PCORE and CS~Dept.\ datasets comprise hundreds of gigabytes of network activity and do not include direct IP-to-user mappings like CPTC. Label creation therefore must depend on other information.

For the CS~Dept.\ datasets, we have access to DNS records for the network. We established ground truth by leveraging the academic network's DNS naming conventions, which provide accurate device identification through systematic naming practices maintained by network administrators.

To create human-predominant device activity labels, 
we leveraged explicit DNS records maintained by network administrators that associate fixed IP addresses with human-operated devices. The academic network's naming convention provides unambiguous identification through exemplar entries like ``johns-laptop.cs.university.edu'' or ``student-macbook-pro.dorm-room-401.university.edu'' that explicitly identify both device type and human ownership.

Direct DNS lookup of all IP addresses observed during each semester enabled semi-automatic labeling. Devices with DNS entries containing human identifiers (personal names, ``laptop,'' ``desktop,'' ``workstation'') or assigned to human-occupied spaces (offices, dorm rooms, lab assignments) were classified as human-predominant. Only fixed IP assignments were included to ensure consistent attribution throughout data collection, excluding dynamic assignments that cannot be reliably tracked across sessions.

To create automation-predominant device activity labels,
the same DNS methodology identified devices with explicit non-human designations, such as ``hvac-thermostat-building3.facilities.university.edu'' or ``gpu-cluster-node15.hpc.university.edu''. These entries indicate automated systems, including network infrastructure, building automation, laboratory equipment, servers, and IoT devices.

DNS records containing functional descriptors (``server,'' ``printer,'' ``sensor,'' ``controller'') or building automation designations were classified as non-human. 

This approach ensured consistent labeling based on documented device functions rather than subjective interpretation for the CS Dept. datasets.

\subsubsection{Labeling PCORE}

For the PCORE datasets, neither direct IP mappings nor DNS information are available. Consequently, we apply a rule-based labeling system (described below) 
to assign human-predominant and automation-predominant labels for training.

We describe here a system for generating labels for datasets that lack ground-truth human vs. not-human,  DNS-verified ground truth, or have other limits that make it impractical to manually label the dataset.   We developed a rule-based classification system that examines connection log characteristics across multiple dimensions to distinguish human-operated from automated devices. It is critical to understand that this classifier serves only to provide binary labels (human-predominant vs. automation-predominant) for training data preparation. The rule-based classifier does not measure behavioral fidelity, degrees of human-likeness, or how convincingly a device exhibits human-like patterns. The latter is precisely PHASE's purpose: to quantitatively assess the realism of network behavior once devices are labeled. The rule-based classifier answers ``what should this device be labeled as?'' while PHASE answers ``how human-like does this device actually behave?''

The classifier analyzes behavioral patterns across six weighted dimensions: DNS query diversity (40\%), HTTP User-Agent patterns (20\%), temporal activity patterns (15\%), protocol diversity (10\%), authentication patterns (10\%), and connection characteristics (5\%). These weights and categories were manually selected by security experts based on observed characteristics of human network behavior and insights from prior work on distinguishing human from automated activity. The rule-based classifier assigns composite scores by evaluating features including domains queried per day, browser User-Agent detection, business hours activity concentration, Kerberos principals, SSH sessions, and SSL server diversity. The framework identifies human indicators (browser User-Agents with diverse browsing, business hours patterns, high work-hours SSH) and automation signatures (24/7 operation, API User-Agents, low DNS diversity, User-Agent rotation, repeated queries to limited domains)~\cite{panza2022extracting, baena2025comprehensive, herrmann2013behavior, bustos2020identifying}. It is important to note that this classifier was developed specifically for labeling the PCORE datasets and is not intended as a universal human detection system. HTTP User-Agent data used by the rule-based classifier was sourced from the PCORE dataset and received in anonymized form; all collection, anonymization, and usage compliance were managed by the third-party dataset providers.

Before applying this classifier to PCORE, we validated the classifier's performance against the CS Dept. datasets where DNS-verified ground truth labels exist. As shown in~\autoref{tab:rulebasedperformance}, the classifier achieved 91.9\% accuracy when validated against known human and non-human devices, with 75.7\% recall for human devices. This validation demonstrates that the rule-based approach produces labels that are reliable enough for  training  the PHASE models with the PCORE dataset. After validation, we applied this classifier to the PCORE datasets to establish their device labels. While validated against CS Dept. datasets with DNS-verified ground truth (\autoref{tab:rulebasedperformance}), the classifier was applied to PCORE datasets where such verification was not available. Consequently, PHASE's ability to distinguish automation-predominant from human-predominant activity on PCORE is inherently limited by the quality of these rule-based labels. Therefore, we treat PHASE performance metrics on PCORE as indicative rather than definitive.

\begin{table}[hbtp]
\scriptsize
\centering
\resizebox{0.7\columnwidth}{!}{%
    \begin{tabular}{lc}
        \toprule
        \textbf{Metric} & \textbf{Rule-Based Classifier}\\
        \midrule
        Accuracy & $0.919$\\
        Balanced Accuracy & $0.850$\\
        Precision & $0.667$\\
        Recall & $0.757$\\
        F1 Score & $0.709$\\
        AUC & $0.876$\\
        \bottomrule
    \end{tabular}
}
\vspace{2 pt}
\caption{Rule-based Classifier Performance on Ground Truth Validation Dataset}
\label{tab:rulebasedperformance}
\end{table}

\subsection{Dataset Statistics}
Statistics from all datasets used to train and evaluate PHASE are shown in~\autoref{tab:datasets}. The eight datasets span four distinct operational environments with different ground truth labeling methodologies: DNS-verified labels from computer science department networks (CS Dept. datasets), rule-based classification from large academic institutional networks (PCORE datasets), and organizer-provided labels from competition network captures (CPTC datasets). 

This evaluation encompasses realistic diversity across multiple dimensions. Scale ranges from 12 to 260 million connection logs. Temporal coverage spans 4 to 125 contiguous days. Network populations vary from small competition scenarios (13-23 devices) to large enterprise deployments (2,780+ devices). Operational contexts include academic research networks, production university infrastructure, and competition simulation environments. This breadth of evaluation provides comprehensive validation of PHASE's capabilities.

\begin{table*}[hbtp]
\centering
\scriptsize
\renewcommand{\arraystretch}{0.85}
    \begin{tabular}{lccccc}
        \toprule
        \textbf{Dataset} 
                         & \textbf{Contiguous Days} & \textbf{Connection Logs} & \textbf{Human IPs} & \textbf{Not Human IPs} \\
                         
        \midrule
        CS Dept. Summer 2024 & 52 & 48.7 Million & 50 & 229 \\
        CS Dept. Fall 2024   & 125 & 140 Million & 103 & 250 \\ 
        CS Dept. Spring 2025 & 116 & 112.4 Million & 117 & 258 \\
        CS Dept. Summer 2025 & 89 & 106.7 Million & 117 & 258 \\
        PCORE 1 Week Small & 8 & 29.1 Million & 332 & 333 \\
        PCORE 1 Week Large & 8 & 260.1 Million & 2780 & 2780 \\
        CPTC 2023: Hotel \& Hospitality & 8 & 43.2 Million & 13 & 19 \\
        CPTC 2024: Air \& Rail & 4 & 11.9 Million & 14 & 23 \\
        \bottomrule
    \end{tabular}
\vspace{2 pt}
\caption{Datasets' Details}
\label{tab:datasets}
\end{table*}

\section{PHASE Model Training Results}
\label{subsection:phase-model-training}

Training results are summarized in~\autoref{tab:modelTrainRes}. Using stratified 10-fold cross-validation, with 1,000 training epochs and a learning rate of 0.001, all PHASE models achieved strong performance in distinguishing between human-predominant and not human-predominant network activity, with most models exceeding 90\% in both accuracy and balanced accuracy. The relatively high standard deviations reflect natural variability in the model's cross-validation performance across different temporal folds. 

\begin{table*}[hbtp]
\centering
\scriptsize
    \begin{tabular}{lcccccc}
        \toprule
        \textbf{Dataset} & \textbf{Accuracy} & \textbf{Bal. Acc.} & \textbf{Precision} & \textbf{Recall} & \textbf{F1} & \textbf{AUC}\\
        \midrule
        CS Dept. Summer 2024 & $.970 \pm .01$ & $.980 \pm .02$ & $.792 \pm .02$ & $.989 \pm .02$ & $.879 \pm .04$ & $.978 \pm .01$ \\
        CS Dept. Fall 2024   & $.939 \pm .07$ & $.938 \pm .08$ & $.905 \pm .10$ & $.935 \pm .09$ & $.919 \pm .09$ & $.938 \pm .08$\\
        CS Dept. Spring 2025 & $.954 \pm .03$ & $.960 \pm .03$ & $.854 \pm .07$ & $.970 \pm .03$ & $.907 \pm .06$ & $.959 \pm .03$\\
        CS Dept. Summer 2025 & $.882 \pm .03$ & $.846 \pm .05$ & $.716 \pm .06$ & $.782 \pm .10$ & $.744 \pm .06$ & $.846 \pm .05$\\
        PCORE 1 Week Small & $.883 \pm .06$ & $.905 \pm .05$ & $.718 \pm .10$ & $.952 \pm .05$ & $.816 \pm .08$ & $.905 \pm .05$\\
        PCORE 1 Week Large & $.965 \pm .04$ & $.968 \pm .04$ & $.911 \pm .08$ & $.977 \pm .03$ & $.942 \pm .06$ & $.968 \pm .04$\\
        CPTC 2023: Hotel \& Hospitality & $.911 \pm .09$ & $.908 \pm .11$ & $.898 \pm .10$ & $.889 \pm .20$ & $.881 \pm .15$ & $.908 \pm .11$\\
        CPTC 2024: Air \& Rail & $.806 \pm .19$ & $.810 \pm .19$ & $.727 \pm .30$ & $.800 \pm .31$ & $.751 \pm .29$ & $.810 \pm .19$\\
        \bottomrule
    \end{tabular}
\vspace{2 pt}
\caption{PHASE Model Performance}
\label{tab:modelTrainRes}
\end{table*}


Among the CS Dept. models, CS Dept. Summer 2024 achieved the highest overall accuracy (0.970) and balanced accuracy (0.980). However, this model exhibited lower precision (0.792), likely due to the reduced presence of human users in academic offices during the summer semester, which may have led to more ambiguous or sparse human activity patterns.

In contrast, CS Dept. Fall 2024 achieved the highest precision (0.905) among all CS Dept. models, despite having lower overall accuracy (0.939). This improvement in precision is likely attributable to the larger and more diverse dataset collected during the fall semester, which enabled the model to better generalize across both human and non-human classes. CS Dept. Spring 2025 offered balanced performance across all metrics, with strong precision (0.854), recall (0.970), and F1 score (0.907). Its AUC of 0.959 further confirms its robustness in distinguishing between classes.

CS Dept. Summer 2025 exhibited notably lower performance across all metrics (accuracy 0.882, precision 0.716, F1 score 0.744), representing the most challenging dataset for classification. Notably, this underperformance differs from the Summer 2024 pattern, where high accuracy but low precision suggested sparse but distinguishable human activity. The CS Dept. network underwent constant changes in network architecture during the Summer 2025 collection period, and additionally experienced multiple extended outages (3-5 days each) due to maintenance and upgrades to the monitoring hardware. These infrastructure disruptions likely degraded classification performance by introducing inconsistencies in data collection and network traffic patterns. The downtime created gaps in behavioral continuity, while architectural changes altered the underlying network characteristics mid-dataset, violating the model's assumption of stable operational conditions. Despite these challenges, PHASE achieved 88.2\% accuracy, demonstrating its robustness and practical utility in real-world operational environments where infrastructure changes and maintenance disruptions are inevitable realities rather than controlled laboratory conditions.

Beyond the CS Dept. datasets, PHASE demonstrated strong generalization across diverse operational environments. The PCORE 1-Week Large model achieved performance comparable to the best CS Dept. models (accuracy 0.965, precision 0.911), despite relying on rule-based rather than DNS-verified labels. This alignment across independently sourced datasets suggests that the rule-based classifier provides labels of sufficient quality for effective PHASE training at scale. While these results should still be interpreted with appropriate caution, given the absence of authoritative ground truth, the consistency in performance indicates that any label noise introduced by the rule-based system did not materially limit PHASE’s ability to learn meaningful behavioral patterns. The CPTC datasets, representing competition environments with multiple simultaneous human actors, showed respectable performance (CPTC 2023 accuracy 0.911, CPTC 2024 accuracy 0.806). The lower accuracy scores for CPTC 2024 are likely due to the issues regarding missing features in the dataset as detailed in~\autoref{subsubsection:CPTC-construction}.

Overall, the F1 scores across all models (ranging from 0.744 to 0.942) highlight the PHASE DNNs' effectiveness in managing the trade-off between precision and recall across diverse environments. The consistently high AUC values (most above 0.90) further validate the models' ability to distinguish between human and non-human activity across different operational contexts and temporal periods. These results answer RQ1: passive network log analysis can reliably distinguish human-predominant from automation-predominant activity across diverse operational environments.





\subsection{PHASE Feature Explanations}
\label{subsection:phase-feature-expl}
While PHASE's high classification accuracy demonstrates its ability to distinguish human from automated network activity, understanding how the model makes these distinctions is essential for improving synthetic user personas. Without interpretability, PHASE would function as a black box, providing scores but no actionable guidance for SUP developers seeking to enhance behavioral realism. To address this need, we apply SHAP (SHapley Additive exPlanations) analysis to reveal which network behaviors and temporal patterns characterize genuine human activity~\cite{shapley}. 


This analysis serves three critical purposes. First, it validates that PHASE learns meaningful behavioral patterns rather than spurious correlations or dataset artifacts. Second, it identifies specific deficiencies in synthetic personas by showing which features deviate from human-like patterns. Third, it provides concrete, data-driven guidance for improving SUP design, as demonstrated in our MCHP case study (\autoref{subsection:sup-eval}).

SHAP quantifies each feature's contribution to the model's predictions across 1,440 time steps (1,440 minutes per day). We analyze SHAP importance from two complementary perspectives: temporal patterns across the day that reveal when different features become most discriminative (\autoref{subsection:SHAP-time-wise}), and feature-specific influences at individual time steps that show how feature values drive classification decisions (\autoref{subsection:SHAP-feature-wise}). All results focus on the PHASE model trained on the Spring 2025 dataset, which is representative of broader trends across the CS Dept. models.


\subsubsection{Temporal Importance Patterns}
\label{subsection:SHAP-time-wise}

\begin{figure*}[hbtp]
  \centering
  \includegraphics[width=\textwidth]{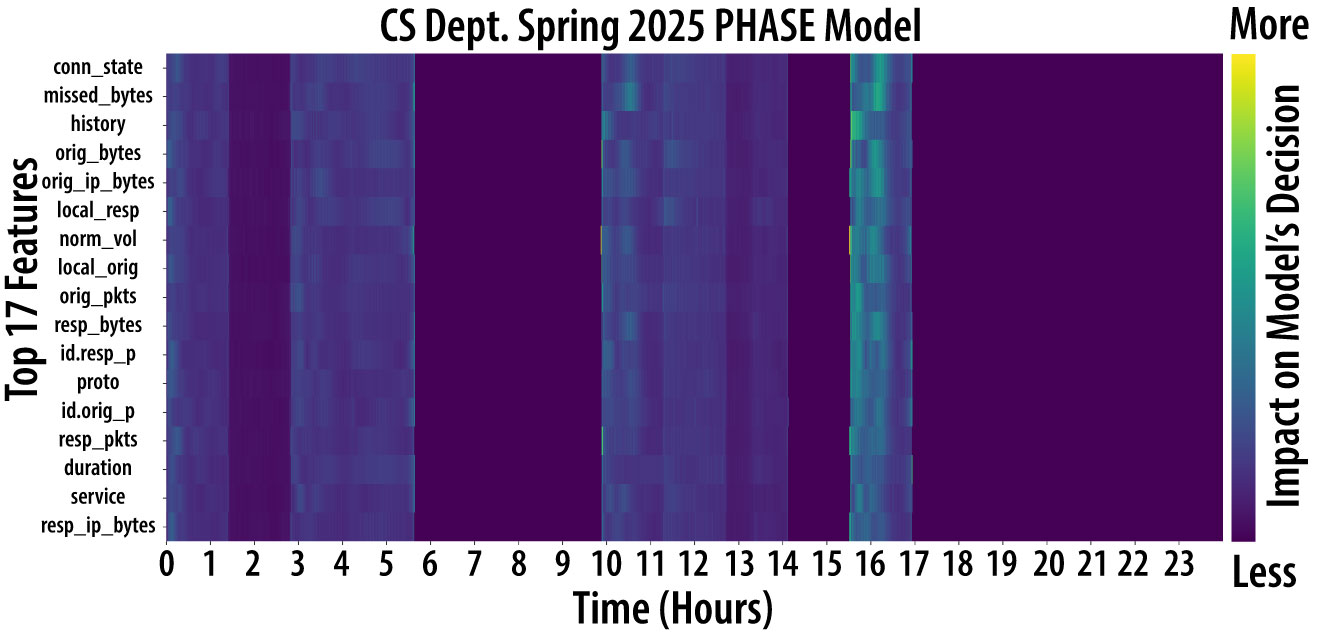}
  \caption{Heatmap of SHAP feature importance at each time step for PHASE Spring 2025 Model}
  \label{fig:topHeatmap}
\end{figure*}

\autoref{fig:topHeatmap} shows feature importance across all time steps, where brighter colors indicate higher absolute SHAP values. The model identifies four distinct periods of heightened importance: 12:01 AM to 1:25 AM, 2:49 AM to 5:39 AM, 9:52 AM to 2:00 PM, and 3:31 PM to 5:00 PM EST. These patterns align with real academic behavior: gradual activity ramp-up after 9:00 AM, peak activity during work hours, and decline after 5:00 PM. Early morning spikes likely correspond to automated system maintenance rather than direct human activity. However, even these automated maintenance tasks contribute to the overall behavioral signature of human-predominant devices. Human workstations typically have regular automated updates, backups, and system scans scheduled during off-hours to avoid disrupting user work, whereas purely automated systems such as servers or IoT devices exhibit different maintenance patterns with more continuous or randomly distributed update schedules. This distinction in maintenance timing becomes part of the discriminative behavioral profile.

Three features emerge as most influential overall: \texttt{conn\_state} (connection termination status), \texttt{missed\_bytes} (unobserved expected bytes), and \texttt{history} (packet-level event sequence). However, every feature becomes most important at least once during the day, indicating the model captures distributed behavioral patterns rather than relying on dominant features.

\begin{figure*}[hbtp]
  \centering
  \includegraphics[width=\textwidth]{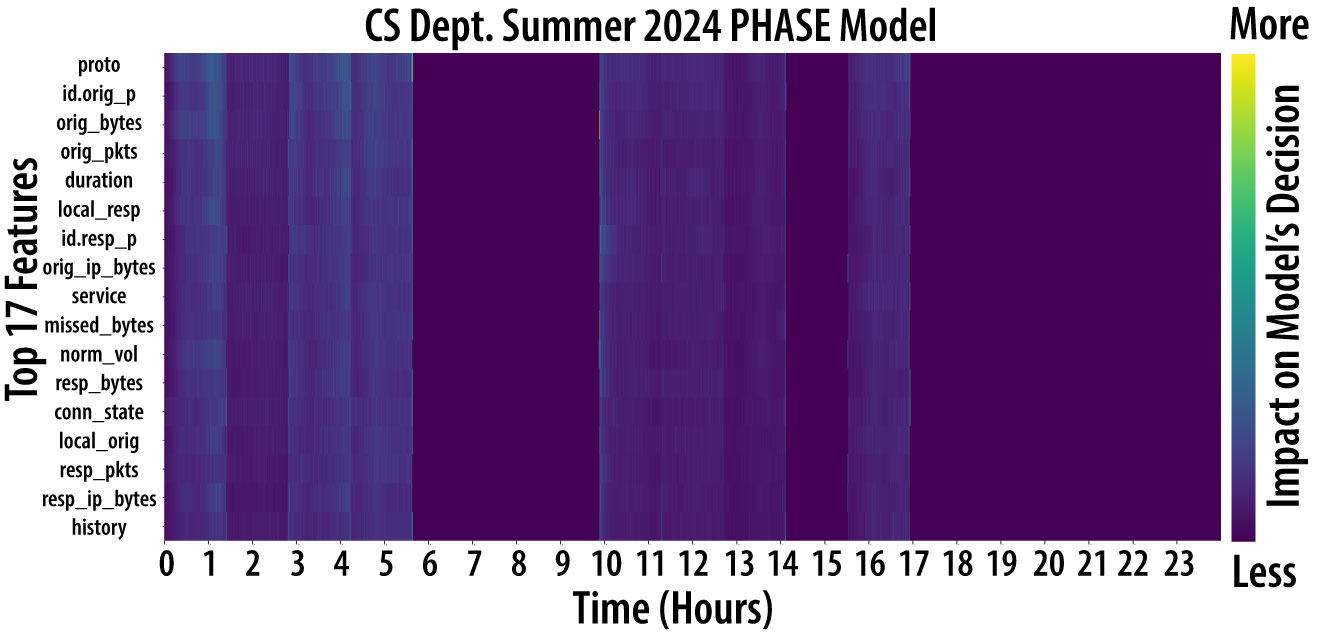}
  \caption{Heatmap of SHAP feature importance at each time step for PHASE Summer 2024 Model}
  \label{fig:topHeatmap29.3}
\end{figure*}

Comparing with the Summer 2024 model (\autoref{fig:topHeatmap29.3}) reveals different temporal patterns and importances despite similar 97\% accuracy. The summer model shows less importance in the evening hours, reflecting flexible academic schedules and fewer users. The SHAP heatmaps suggest that PHASE may have some adaptability to different network contexts, leveraging temporal patterns when available or intrinsic feature characteristics when temporal structure is less pronounced.

Feature importance varies across models trained on different datasets. The Spring model prioritizes the history string, which captures packet-level connection sequences, while the Summer model focuses more heavily on protocol types and \texttt{orig\_bytes}. This suggests that during summer months, when fewer users are present and schedules are more flexible, the fundamental characteristics of host connections become more discriminative for identifying human activity than detailed interaction patterns. These changes in PHASE's identification patterns indicate how context-specific human behavior can be across different temporal contexts. 

\subsubsection{Feature-Specific Analysis}
\label{subsection:SHAP-feature-wise}


SHAP beeswarm plots reveal how individual feature values influence predictions at specific time steps. Points to the right (positive SHAP values) indicate features that push the model toward classifying the activity as human-like, while points to the left (negative SHAP values) push toward non-human classification. The color gradient represents feature values from low (dark purple) to high (light yellow), showing how the magnitude of each feature affects its contribution to the prediction.

\begin{figure}[hbtp]
  \includegraphics[width=\columnwidth]{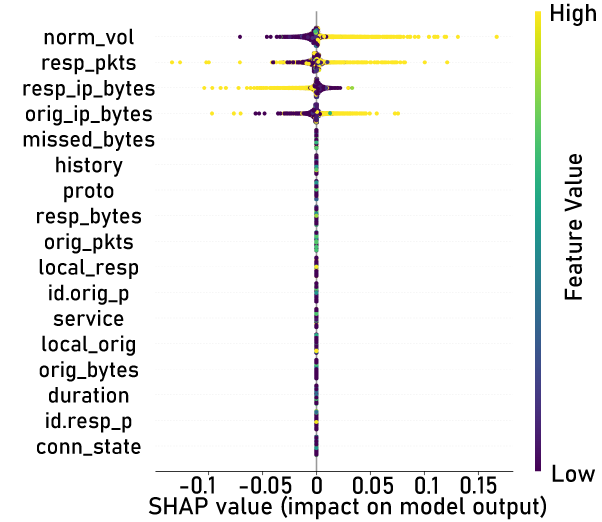}
  \caption{Time step 931 (3:31 PM EST) SHAP beeswarm feature importance}
  \label{fig:bee931}
\end{figure}

At time step 931 (3:31 PM, ~\autoref{fig:bee931}), \texttt{norm\_vol} and \texttt{resp\_pkts} are most influential. High values of both features strongly indicate human activity. 

\begin{figure}[hbtp]
  \includegraphics[width=\columnwidth]{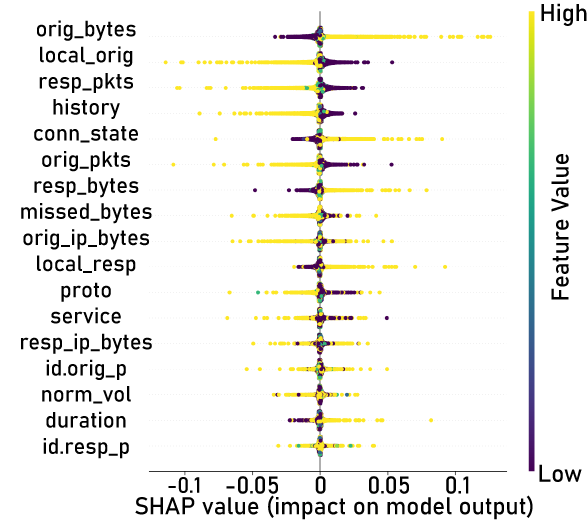}
  \caption{Time step 593 (9:53 AM EST) SHAP beeswarm feature importance}
  \label{fig:bee593}
\end{figure}

Feature importance shifts dramatically throughout the day. At time step 593 (9:53 AM, \autoref{fig:bee593}), \texttt{orig\_bytes} and \texttt{local\_orig} become most important, demonstrating the model's sensitivity to different behavioral patterns at different times.

The SHAP analysis reveals distinct behavioral patterns that evolve throughout the day. Early morning activity (e.g., time step 593) is characterized by high \texttt{orig\_bytes} and \texttt{local\_orig} values being false, indicating human activity involves substantial outbound data transfer to external destinations. This pattern suggests users initiating connections and sending significant data volumes outside the local network.

Later in the day (e.g., time step 931), the signature shifts to high connection volumes \texttt{norm\_vol} and response packets \texttt{resp\_pkts} but relatively low \texttt{resp\_ip\_bytes} combined with high \texttt{orig\_ip\_bytes}. This asymmetric pattern, many outbound connections with limited inbound data, aligns with typical web browsing behavior, where users initiate numerous requests but receive relatively small response payloads. These temporal variations demonstrate how PHASE captures the natural rhythm of human network activity, from data-intensive morning tasks to interactive browsing patterns during peak hours. These findings answer RQ2: temporal and behavioral feature signatures indicative of human activity vary substantially across operational contexts, with temporal structure itself determining which features are most discriminative.

\section{Synthetic User Persona Case Study}
\label{section:sup-methodology}
To demonstrate PHASE's practical utility for improving synthetic user personas, we conducted a case study evaluating the widely deployed MITRE Caldera Human Plugin (MCHP)~\cite{alford2022caldera} with the CS Dept. models trained in the Summer 2024, Fall 2024, and Spring 2025 semesters. This evaluation applies PHASE's temporal and behavioral insights from~\autoref{subsection:phase-feature-expl} to identify specific deficiencies in synthetic behavior and develop data-driven improvements. We present the deployment methodology (\autoref{subsection:SUP-deployment}), PHASE evaluation results (\autoref{subsection:sup-eval}), and an enhanced configuration that achieves measurably more human-like behavior (\autoref{subsubsection:dopey-eval}).

\subsection{MCHP Deployment and Configuration}
\label{subsection:SUP-deployment}
We selected MCHP for evaluation due to its widespread adoption in large research efforts~\cite{CASTLE, CHASE, math11163448}. MCHP provides customizable workflows that simulate common user actions, including web browsing and YouTube access through the Selenium WebDriver framework~\cite{selenium}, as well as shell command execution.

MCHP was deployed on 14 Ubuntu Server 23.10 virtual machines within an isolated server cluster using fixed IP addresses for consistent activity tracking. We removed Windows-specific workflows and updated Chrome browser components to ensure compatibility. The evaluation used MCHP's default configuration, including predefined sleep intervals, workflow frequencies, and task sequences, to establish a baseline representing MITRE's standard approach to realistic human behavior simulation.

This default configuration serves as a representative example of current SUP technology deployed in operational cybersecurity simulations, making improvements to its behavioral fidelity immediately applicable across numerous existing deployments.

\subsection{MCHP Evaluation}
\label{subsection:sup-eval}

For evaluating the MCHP behavioral fidelity, we pick three datasets from the CS~Dept.\ datasets, and their corresponding trained PHASE models.  
Each model was used to classify daily MCHP activity embedded in the dataset, producing probability scores (0-1) that quantify how closely synthetic behavior resembles the real human in that dataset.
We ensured that MCHP data was embedded in our datasets by running MCHP for the Summer 2024, Fall 2024, and Spring 2025 semesters. The Summer 2025 semester was excluded from MCHP evaluation because the server cluster experienced extended downtime during this period, preventing consistent MCHP data collection. Similarly, we could not run MCHP for the external datasets from CPTC and PCORE.  As it would be unfair to MCHP to evaluate it on a network topology that is different than where it is being evaluated, we exclude these datasets from evaluation.  We evaluate both the default MCHP configuration (\autoref{subsubsection:default-eval}) and an enhanced version based on PHASE insights (\autoref{subsubsection:dopey-eval}).

To interpret PHASE scores meaningfully, we established thresholds based on the lowest observed model accuracy for the CS Dept. datasets (88.2\%) and its standard deviation ($\pm$3\%). We set the confidently human-like threshold at 0.8, representing approximately one standard deviation below the lowest observed model accuracy (0.882 $-$ 0.03 $=$ 0.852), rounded down to the nearest 0.2-width interval for consistency with the gradations described below.
The remaining score ranges were divided into equal 0.2-width intervals to create consistent gradations of confidence: $1 \geq S \geq 0.8$ indicates confidently human-like behavior within one standard deviation of model accuracy; $0.8 > S \geq 0.6$ indicates likely human activity above the classification threshold but with lower confidence; $0.6 > S \geq 0.4$ is ambiguous with no clear classification; $0.4 > S \geq 0.2$ suggests likely non-human behavior closer to automated activity; and $0.2 > S \geq 0$ indicates confidently non-human activity strongly aligned with automated classification.

\begin{figure}[hbtp]
\centering
  \includegraphics[width=\columnwidth]{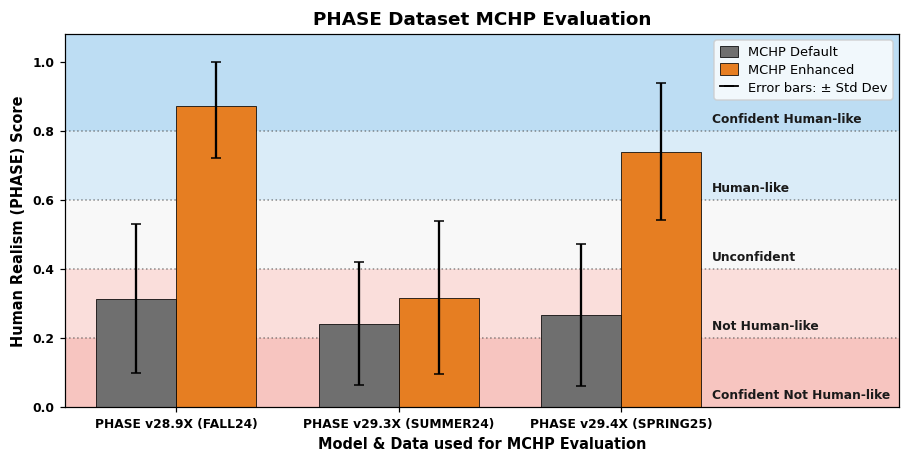}
  \caption{MCHP evaluation across three different models trained on the three CS DEPT. Datasets}
  \label{fig:mchpeval}
\end{figure}

These thresholds enable systematic comparison of SUP configurations and identification of specific behavioral deficiencies. The evaluation results across all models are presented in~\autoref{fig:mchpeval}, demonstrating PHASE's ability to provide actionable feedback for SUP improvement rather than binary classification alone.

\subsubsection{Default MCHP Evaluation}
\label{subsubsection:default-eval}
As illustrated in~\autoref{fig:mchpeval}, default MCHP consistently scores below the human-like behavior threshold across all models. The Summer 2024 PHASE model assigned the lowest mean score of $0.24 \pm 0.18$, while the Fall 2024 PHASE model yielded the highest at $0.31 \pm 0.22$, still well below the 0.5 classification boundary. The large standard deviations indicate the variation in different days of activity, though variability remains within the non-human activity range. 

\begin{figure}[hbtp]
\centering
  \includegraphics[width=\columnwidth]{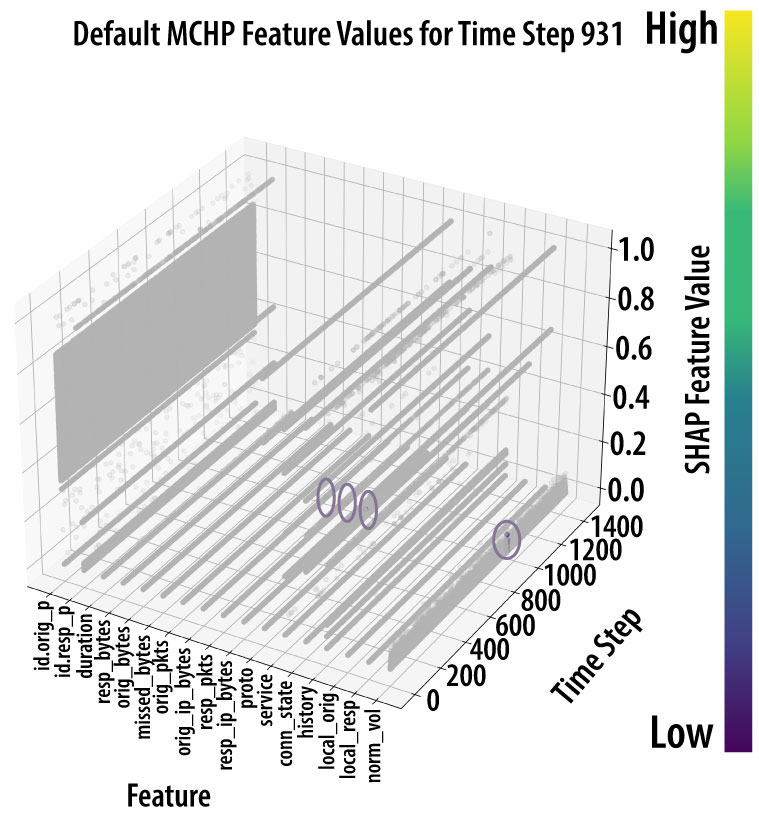}
  \caption{Time step 931 (3:31 PM EST) SHAP 3D graph of default MCHP feature values}
  \label{fig:mchpshap931}
\end{figure}

To understand why PHASE consistently classifies MCHP as non-human, we applied SHAP analysis to examine the model's decision-making process. \autoref{fig:mchpshap931} shows feature values at time step 931 (3:31 PM EST), a period our earlier analysis identified as behaviorally significant for distinguishing human activity.

The analysis reveals that MCHP exhibits low values for the most discriminative features: \texttt{norm\_vol} (connection volume), \texttt{resp\_pkts} (response packets), and \texttt{orig\_ip\_bytes} (outbound data volume). These low values create negative SHAP contributions, driving the non-human classification of the MCHP plugin's activity. 

This pattern suggests three fundamental behavioral deficiencies in the default MCHP configuration: low \texttt{norm\_vol} indicates insufficient connection diversity compared to typical human users who maintain multiple simultaneous applications; low \texttt{resp\_pkts} suggests limited interactive engagement, with MCHP's browsing involving simple page requests rather than dynamic content loading; and low \texttt{orig\_ip\_bytes} reflects reduced data consumption relative to human users who generate substantial outbound traffic through uploads, video calls, and application usage.

These deficiencies reflect MCHP's focus on basic automation tasks rather than the rich, multifaceted network behavior characteristic of real human computer usage. The low scores across multiple time steps indicate these limitations persist throughout the day, creating an identifiable synthetic signature that distinguishes MCHP activity from authentic human behavior. These findings answer RQ3: default MCHP activity is behaviorally distinguishable from real human users, consistently scoring in the non-human range across all evaluated models.

\subsubsection{Enhanced MCHP Evaluation}
\label{subsubsection:dopey-eval}
Based on SHAP insights revealing MCHP's deficient connection patterns, we developed the MCHP Enhanced configuration with a strategic behavioral modification: periodic one-hour idle periods after completing randomized task sequences. This enhanced approach leverages temporal camouflage to better disguise synthetic activity within natural human behavioral rhythms. The one-hour duration was selected to align with typical human meeting lengths and lunch breaks observed in the training data.

As shown in~\autoref{fig:mchpeval}, MCHP Enhanced achieves consistently higher PHASE scores across all models without fundamentally altering the underlying connection characteristics. The MCHP Enhanced behavior scores 55 percentage points more human-like than the MCHP Default behavior in the Fall 2024 PHASE evaluation (0.86 vs 0.31). The MCHP Enhanced behavior scores 47 percentage points more human-like than the MCHP Default behavior in the Spring 2025 PHASE evaluation (0.71 vs 0.24). Finally, the MCHP Enhanced behavior scores 7 percentage points more human-like than the MCHP Default behavior in the Summer 2024 PHASE evaluation (0.31 vs 0.24). This improvement demonstrates that when activity occurs is critically important for behavioral fidelity.

The periodic rest strategy leverages a key insight from our temporal analysis (\autoref{subsection:SHAP-time-wise}): human network behavior exhibits natural gaps corresponding to meetings, breaks, and task transitions. By introducing structured idle periods, MCHP Enhanced creates temporal signatures that align with expected human patterns, effectively masking its limited connection diversity within realistic activity windows.
\begin{figure}[h]
\centering
  \includegraphics[width=\columnwidth]{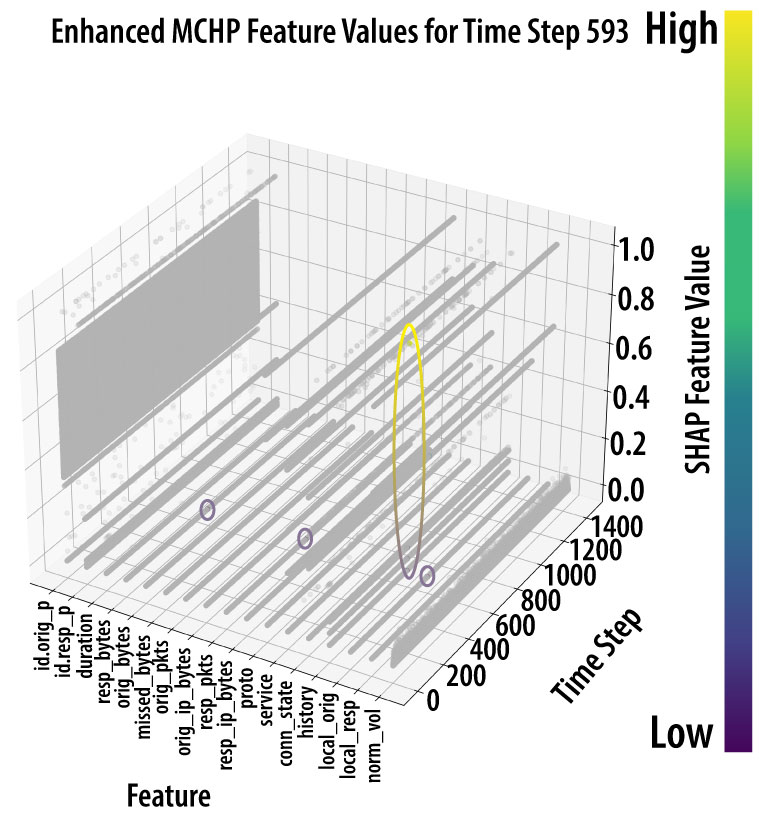}
  \caption{Time step 593 (9:53 AM EST) SHAP 3D graph of enhanced MCHP feature values}
  \label{fig:mchpshap593}
\end{figure}

The SHAP analysis at time step 593 (\autoref{fig:mchpshap593}) reveals how temporal camouflage affects feature interpretation. While \texttt{orig\_bytes} remains low (typically indicating non-human activity), the combination with low values for \texttt{local\_orig}, \texttt{resp\_pkts}, and \texttt{history} now contributes positively toward human classification. This suggests that within the context of expected human idle periods, reduced activity levels become consistent with authentic behavior rather than synthetic limitations.

This behavioral camouflage effect explains why MCHP Enhanced scores higher on the Fall 2024 and Spring 2025 models, without addressing the fundamental connection volume deficiencies identified in our earlier analysis (\autoref{subsection:phase-feature-expl}). The idle periods create temporal contexts where limited network activity appears natural, reducing the detectability of MCHP's synthetic signatures.

However, the temporal camouflage strategy reveals important limitations when examined across different seasonal contexts. The Summer 2024 PHASE model still classifies MCHP Enhanced as non-human-like, showing only a 7 percentage point improvement and highlighting a critical weakness in this approach.

Our earlier analysis (\autoref{subsection:SHAP-time-wise}) showed that the Summer model exhibits reduced temporal structure, with feature importance distributed more uniformly across time steps rather than concentrated in specific periods. This reflects the reality of academic summer environments where work schedules are flexible, fewer users are present, and traditional temporal rhythms are less pronounced. Consequently, the Summer model places greater emphasis on intrinsic activity characteristics: connection volumes, data patterns, and interaction types, rather than temporal alignment.

This emphasis on intrinsic activity characteristics over temporal alignment explains why temporal camouflage fails for the Summer model: when human behavior itself lacks rigid temporal structure, idle periods become less meaningful for classification. The Summer model's focus on activity quality over timing exposes MCHP's fundamental deficiencies in connection diversity and data exchange patterns that periodic rests cannot mask. These findings answer RQ4: data-driven insights derived from behavioral analysis can measurably improve the humanness of synthetic user personas, though the degree of improvement is highly dependent on operational context.



\subsubsection{Implications for SUP Design}
This MCHP evaluation demonstrates four critical insights for synthetic persona development. First, temporal camouflage strategies can dramatically improve behavioral fidelity, achieving up to 55 percentage point improvements in structured environments, without modifying fundamental activity characteristics. Second, behavioral authenticity is highly contextually dependent: strategies effective during structured academic semesters (Fall/Spring) largely fail during periods with fluid schedules (Summer), revealing that no single improvement approach generalizes across all operational contexts. Third, without quantitative measurement frameworks like PHASE, developers cannot identify these context-specific vulnerabilities or distinguish between apparent improvements that introduce new detectability vectors versus genuine enhancements in behavioral realism. Fourth, PHASE measures complete device-level behavioral profiles rather than isolated human-initiated actions, as background automation on human workstations is an unavoidable component of real network activity that SUPs must replicate. Consequently, PHASE scores are relative to a specific operational environment rather than a universal realism threshold.

These findings underscore PHASE's essential role in SUP development: providing systematic, data-driven measurement that enables developers to validate behavioral fidelity across diverse operational environments rather than relying on subjective assessment or single-context evaluation. The framework's SHAP-based interpretability further enables targeted improvements by revealing specific temporal and behavioral deficiencies, as demonstrated by the enhanced MCHP configuration developed from PHASE insights. This evaluation utilized three CS Dept. datasets (Summer 2024, Fall 2024, Spring 2025); the Summer 2025 dataset was excluded due to extended infrastructure downtime during collection.

\section{Related Work}
\label{section:related-work}
While research on cyber ranges, honeypots, and sandboxes has expanded, most work focuses on attack-defense scenarios rather than the fidelity of simulated network behavior~\cite{s20247148}. Only a small fraction of cyber ranges (approximately 8\%) prioritize realistic traffic simulation, revealing a significant gap in the literature~\cite{s20247148}. Some approaches generate synthetic activity using techniques like dynamic word embeddings and spatiotemporal modeling, or by blending real and simulated traffic to augment datasets~\cite{ring2017flow,10048962}. Others, such as CybORG and Microsoft's CyberBattleSim, simulate contextualized agent actions using reinforcement learning frameworks~\cite{baillie2020cyborg, msft:cyberbattlesim}. However, these systems often assume human-like behavior without quantitatively validating it against real-world network activity, limiting their ability to model authentic user behavior. 

Long Short-Term Memory (LSTM) networks are widely used in cybersecurity, particularly for Network Intrusion Detection Systems (NIDS), where hybrid LSTM architectures have shown strong performance on public datasets~\cite{10101759}. More recently, foundational models such as LENS have demonstrated the potential of large-scale, adaptable architectures for general traffic prediction using raw PCAP data~\cite{wang2024lens}. Despite their effectiveness, LSTMs and other deep learning models are often criticized for their lack of interpretability, a significant concern in cybersecurity where transparency is essential for adoption~\cite{apruzzese2023sok, 9527908}. As a result, explainability remains a critical requirement for deploying machine learning solutions in the cybersecurity domain. 

Human Activity Recognition (HAR), a well-established field outside cybersecurity, demonstrates the effectiveness of LSTM-based architectures for classifying behavioral patterns from temporal data. Hassan et al. employed MobileNetV2 combined with Deep BiLSTMs to extract and classify subtle visual cues such as hats or other features from image sequences~\cite{app14020603}. Other researchers have applied LSTMs to detect behavioral anomalies in collaborative tasks and transformed passive RFID signals into meaningful behavioral classifications using hybrid transformer models~\cite{10.1145/3568294.3580208, LIU2023}. 

These HAR approaches share a common insight with PHASE: human behavior exhibits temporal dependencies that sequential models can effectively capture with minimal preprocessing. While HAR typically analyzes visual or sensor data, the underlying principle applies directly to network connection logs: human activity creates identifiable patterns across time regardless of the data source. PHASE adapts these time series modeling techniques to cybersecurity contexts, treating network telemetry as behavioral sequences that reveal human versus automated activity patterns.

\section{Future Work}
\label{section:future-work-ext}
Currently, PHASE requires labeled datasets for training, limiting deployment to environments where ground-truth human/non-human labels can be established. To expand PHASE's applicability, future work will develop semi-supervised and unsupervised learning approaches that leverage publicly available network datasets without explicit behavioral labels. These advances would enable PHASE deployment across diverse organizational contexts, allowing the growing community of SUP developers to validate behavioral fidelity regardless of their specific network environment or labeling capabilities.

Our results across eight diverse datasets spanning competition networks, academic departments, and large institutional infrastructure demonstrate that PHASE's technique for capturing temporal and behavioral signatures consistently achieves high accuracy across varied operational contexts. Systematically evaluating whether PHASE models trained on one network environment can effectively classify activity in fundamentally different contexts without retraining represents important future work that the scope of this paper precluded.

The landscape of synthetic user personas continues to evolve rapidly, with AI-enabled agent platforms demonstrating varying degrees of human-like behavioral authenticity~\cite{lu2024omniparserpurevisionbased, openai2025operator, huggingface2024smolagents, amazon2025novaact, updyke2018ghosts}. As these systems become more sophisticated, systematic evaluation of their behavioral fidelity across diverse operational contexts becomes increasingly critical. Future work will apply PHASE to evaluate contemporary synthetic persona implementations, including AI-powered agents and autonomous systems, to establish benchmarks for cross-context behavioral authenticity and identify environment-specific behavioral deficiencies that require targeted improvements.

\section{Ethical Considerations}
\label{section:ethical-considerations}

Data collection was reviewed by the academic institution's Institutional Review Board for the Social and Behavioral Sciences (IRB-SBS) and independently by a U.S. Department of Defense Human Research Protections Official (HRPO). Both bodies formally determined that this research does not meet the federal definition of human subjects research, as the pipeline operates entirely on de-identified data with no personal information in any collected field. Full correspondence from both reviewing bodies is retained on file. All network users operated under an institutional acceptable-use policy agreed to upon receiving institutional credentials, which explicitly notifies users that network traffic may be monitored for research and security purposes. 

PHASE collects only Zeek connection log metadata, not packet contents, meaning no browsing data, message content, or personal communications are accessible at any stage. The specific features utilized are shown in~\autoref{tab:connection_log}; connection contents were never recorded or utilized. Raw traffic is processed in real time by Zeek sensors, which immediately discard underlying packet data. At no stage did this research involve analysis of unencrypted traffic content: Zeek connection logs contain only metadata and never capture payload regardless of underlying encryption, and all User-Agent data used in the rule-based classifier was received in pre-anonymized form from third-party dataset providers. IP addresses, the only personally identifiable information collected, are cryptographically anonymized using LibCryptoPANT~\cite{cryptopant} and further reduced to arbitrary numeric identifiers with no correspondence to any real address. Non-anonymized intermediate files are deleted immediately after anonymization, and no data was transferred off the storage cluster at any point during collection, training, or analysis. Neither the rule-based classifier nor PHASE ever operated on de-anonymized information at any stage. Critically, PHASE classifies activity at the device level only and makes no attempt to trace, profile, or re-identify any individual user or IP address.

\section{Conclusion}
\label{section:conclusion}
Cybersecurity simulation environments, such as cyber ranges, honeypots, and sandboxes, require realistic human activity to be effective, yet no standardized framework existed to measure the behavioral fidelity of the synthetic user personas that generate it. To address this gap, we developed PHASE. Through comprehensive evaluation across diverse operational contexts, we demonstrated the following key contributions: A novel measurement and analysis framework that quantifies synthetic user persona behavioral fidelity by classifying network activity patterns as human-predominant or automation-predominant. PHASE's machine learning framework achieved on average 91\% accuracy in network activity classification across 8 distinct network datasets. A SHAP-based interpretability analysis that revealed specific temporal and behavioral signatures indicative of genuine human network activity, demonstrating that \textbf{when} activity occurs is critically important for behavioral fidelity. Using these insights, we improved the widely used MITRE Caldera Human Plugin's realism by up to 55 percentage points across three distinct datasets through temporal camouflage strategies. A novel data processing pipeline to transform raw connection logs for time series machine learning analysis. Four temporally distinct datasets collected on an operational network, with 50–117 human IP addresses and 229–258 non-human IP addresses per semester. Each dataset comprises over 150 GB of raw connection logs, spanning multiple months of continuous data.

Our evaluation revealed a critical finding about synthetic persona design: human behavior is highly contextually variable. While temporal camouflage strategies achieved 55 percentage point improvements during structured academic semesters, the same approach yielded only 7 percentage point improvements during summer periods with fluid schedules. This demonstrates that effective synthetic persona design requires understanding how behavioral patterns fundamentally shift across different operational contexts, not just general human behavior patterns.

These results establish that PHASE provides essential measurement capabilities for advancing cyber simulation environments beyond assumption-based persona development. Without quantitative evaluation frameworks like PHASE, researchers cannot identify when apparent improvements in synthetic behavior introduce new detectability vectors or when intuitively realistic strategies diverge from authentic human activity patterns. As synthetic user personas become increasingly sophisticated and widely deployed in cybersecurity simulations, PHASE enables the systematic validation necessary to ensure these environments produce authentic network behaviors that support effective cybersecurity training, research, and defense strategy development.

\bibliographystyle{IEEEtran}
\bibliography{bibtex}

\clearpage

\section*{Appendix}
\renewcommand{\thesection}{\Alph{section}}
\setcounter{section}{0}

\appendix
\renewcommand{\thesection}{\Alph{section}}
\setcounter{section}{0}

\section{Acknowledgments}
\label{section:ack}
This material is based on research sponsored by DARPA under agreement number HR001124C0425. The U.S. Government is authorized to reproduce and distribute reprints for Governmental purposes notwithstanding any copyright notation thereon. The views and conclusions contained herein are those of the authors and should not be interpreted as necessarily representing the official policies or endorsements, either expressed or implied, of DARPA or the U.S. Government. The authors thank the University of Virginia computer science department for providing the network infrastructure and operational support that made this research possible, and the members of our research group for their contributions throughout data collection, system development, and evaluation.

\section{Data Availability}
\label{section:data-avail}
The PHASE framework and trained models are publicly available at \url{https://github.com/LampSteven17/PHASE}. The raw network datasets collected as part of this research contain sensitive operational network traffic and are not publicly released. Researchers seeking access to the datasets may contact the authors directly; access will be granted on a case-by-case basis subject to a data use agreement and approval by the relevant institutional review bodies.

\section{Use of Generative AI}
\label{app:useofgenai}
Generative AI was used in this research to help revise the academic paper for typos, errors, clarity, sentence and paragraph refinement, and to debug errors and issues in code. Additionally, generative AI was used to help format citations into BibTeX format for sources that did not provide ready-made BibTeX citations. 

However, all figures and graphs in this paper were hand-generated or coded using Python scripts. For publishing purposes, titles and legends were updated with the use of editing tools, but the results themselves were not changed or altered, by humans or AI, in any way. 

The models used in this research include Claude Sonnet, Claude Opus, Microsoft Copilot, and ChatGPT~\cite{claude_sonnet_2025, claude_opus_2025, microsoft_copilot_2026, openai2024chatgpt}. The use of Grammarly was also included for further refinement~\cite{grammarly}.

\section{Verifying Absence of Data Leakage}
\label{app:data-leakage}
To verify the integrity of the PHASE pipeline and confirm the absence of data leakage, a random dataset was created consisting of 50 human and 50 non-human randomly generated IP addresses with shuffled activity patterns. Since this data contains no genuine signal, any model accuracy substantially exceeding 50\% would indicate information leakage during preprocessing or training. As shown in~\autoref{tab:dataleakagetest}, PHASE achieved 53.0\% accuracy on the random dataset, showing chance performance and confirming that the pipeline does not leak information. PHASE's performance on the random dataset also shows that the high accuracy observed on real-world datasets (\autoref{tab:modelTrainRes}) reflects genuine learned patterns rather than artifacts of the training process.

\begin{table}[htb]
\centering
\normalsize
    \begin{tabular}{lc}
        \toprule
        \textbf{Metric} & \textbf{Random Dataset}\\
        \midrule
        Accuracy & $0.530 \pm 0.04$\\
        Balanced Accuracy & $0.530 \pm 0.04$\\
        Precision & $0.590 \pm 0.14$\\
        Recall & $0.577 \pm 0.32$\\
        F1 Score & $0.498 \pm 0.19$\\
        AUC & $0.530 \pm 0.04$\\
        \bottomrule
    \end{tabular}
\vspace{2 pt}
\caption{PHASE Data Leakage Test}
\label{tab:dataleakagetest}
\end{table}

\section{Additional Figures \& Tables}
\autoref{app:931FV} shows that low response packet counts drive non-human classification, while high counts (approaching the maximum of 247,192,000) suggest human activity.
\begin{table}[hbtp]
\centering
\resizebox{.6\columnwidth}{!}{%
    \begin{tabular}{ccc}
        \toprule
        \textbf{Statistic} & \textbf{Real Value} & \textbf{SHAP Value} \\
        \midrule
        Min & -1 & 0 \\
        Median & 0 & $\approx$ 0 \\
        Mean & 369 & $\approx$ 0 \\
        Max & 247,192,000 & 1 \\
        \bottomrule
    \end{tabular}
}
\vspace{1pt}
\caption{Feature value translations for resp\_pkts at time step 931}
\label{app:931FV}
\end{table}


\end{document}